\documentclass[a4paper]{article}
\usepackage[margin=25mm]{geometry}
\usepackage{amsmath}
\usepackage{amsfonts}
\usepackage{amssymb}
\usepackage{amsthm}
\usepackage[dvips]{graphicx}
\usepackage{authblk}
\usepackage{cite}

\providecommand{\keywords}[1]
{
  \small	
  \textbf{Keywords:} #1
}

\newcommand{\be}{\begin{equation}}
\newcommand{\eee}{\end{equation}}

\begin{document}

\title{Exact solution of the position-dependent mass Schr\"odinger equation with the completely positive oscillator-shaped quantum well potential}

\author{E.I. Jafarov\thanks{Corresponding author: ejafarov@physics.science.az}}
\author{S.M. Nagiyev\thanks{sh.nagiyev@physics.science.az}}

\affil{Institute of Physics, State Agency for Science and Higher Education\\Javid av. 131, AZ1143, Baku, Azerbaijan}

\date{} 

\maketitle

\begin{abstract}
Two exactly-solvable confined models of the completely positive oscillator-shaped quantum well are proposed. Exact solutions of the position-dependent mass Schr\"odinger equation corresponding to the proposed quantum well potentials are presented. It is shown that the discrete energy spectrum expressions of both models depend on certain positive confinement parameters. The spectrum exhibits positive equidistant behavior for the model confined only with one infinitely high wall and non-equidistant behavior for the model confined with the infinitely high wall from both sides. Wavefunctions of the stationary states of the models under construction are expressed through the Laguerre and Jacobi polynomials. In general, the Jacobi polynomials appearing in wavefunctions depend on parameters $a$ and $b$, but the Laguerre polynomials depend only on the parameter $a$. Some limits and special cases of the constructed models are discussed.
\end{abstract}


\section{Introduction}

The study of the harmonic oscillator problem is of great importance in both classical and quantum mechanics due to the problem plays a flagship role in the explanation of a huge number of physical phenomena. The harmonic oscillator problem in one dimension of the macroscopic sizes can be exactly solved within Newtonian mechanics. A block with mass vibrating on a spring or a simple pendulum is the best example of it. As a result of the well-known solution, one observes that the classical system moves about the equilibrium position and has two turning points, at which its kinetic energy becomes zero and the object starts to recover its initial equilibrium position. In other words, a confined motion that the classical harmonic oscillator exhibits within Newtonian mechanics can be considered one of its main features~\cite{bloch1997}.

The harmonic oscillator problem of the sub-micron sizes within the non-relativistic quantum mechanics drastically differs from similar classical mechanics problem~\cite{landau1991,moshinsky1996}. First of all, one needs to take into account that within the quantum theory the momentum and position operators don't commute~\cite{flugge1971,ohnuki1982}. Therefore, we are dealing with probabilities instead of precise joint measurements of position and momentum. The appearance of the probabilistic nature of the observations requires that the obtained expressions of the distributions will be correct only if they are valid within the whole range of the position or momentum from $-\infty$ to $+\infty$. This solution of the quantum harmonic oscillator is well known. It has a discrete energy spectrum with equidistant energy levels, its wavefunctions of the stationary states are analytically expressed through the Hermite polynomials and vanish at $x=\pm \infty$. This does not mean that it is impossible to consider the quantum oscillator problem that is confined similar to its classical analog. Such a problem also is studied intensively due to its necessity in the explanation of many phenomena mainly coming from nanotechnologies. The quantum approach for such a problem consists of coupling together two of its exactly-solvable problems, namely the quantum harmonic oscillator and the infinite potential well problems. In general, these two coupled problems don't lead to the exact solution of the corresponding Schr\"odinger equation. As we are aware,~\cite{auluck1941} can be considered as a first discussion of the problem, where, energy levels of the confined quantum harmonic oscillator are computed approximately with further application to the proton-deuteron transformation as a source of energy in dense stars. Some exactly soluble quantum harmonic oscillator models are known for the case, if one replaces continuous position and momentum representations with their finite-discrete analogues~\cite{atakishiyev2001,jafarov2011a,jafarov2011b,jafarov2012a,jafarov2012b,jafarov2013}. However, the problem still can be exactly soluble in the continuous configuration representation, if one assumes that the mass of the quantum system is not constant, but varies with position~\cite{mathews1975,schmidt2007,amir2014,quesne2015,karthiga2017,naeim2017,jafarov2020a,jafarov2020b,jafarov2020c,jafarov2021a,jafarov2021b,nagiyev2022a,nagiyev2022b,jafarov2022d}.

We are going to discuss the exact solution of the confinement model of the non-rectangular quantum well, a profile of which consists of the harmonic oscillator potential valid within the finite or semi-infinite positive range. Recently, an exact quantum-mechanical solution for the one-dimensional harmonic oscillator model asymmetrically confined into the infinite well was introduced~\cite{jafarov2022}. Its wavefunctions of the stationary states are expressed through the Jacobi polynomials. Unlike the model with the approximate solutions from~\cite{consortini1976}, the exact solution in~\cite{jafarov2022} is achieved thanks to the replacement of the constant mass $m_0$ with the mass $M\left( x \right)$ specifically varying with the position. It also generalizes the semiconfinement oscillator model with the wavefunctions expressed through the Laguerre polynomials~\cite{jafarov2021,jafarov2022b,jafarov2022c}. Then, we decided that one also can successfully generate the confinement model of a completely positive oscillator-shaped quantum well by using the less-studied orthogonality properties of the Jacobi polynomials. The main feature of the model under construction will be the exhibition of the confinement effect at infinitely positive values of the potential. In other words, the quantum system under consideration will be in the cavity between positive $a$ and $b$ values of position with initial condition $a<b$. Also, the special case can be considered here, when the wall located at position $x=b$ will be driven to the $+\infty$.

We structured the paper as follows: Section 2 presents some basic information about the Jacobi polynomials and generalization of their orthogonality relation from $\left( -1;1\right)$ to $\left( a;b\right)$ with the condition $a<x<b$ being satisfied. Here, also some less-known properties of the Laguerre polynomials and the connection of these properties with the Jacobi polynomials are discussed. Further, Section~3 consists of the main result of the current paper -- exact solutions to the confinement model of the non-relativistic one-dimensional completely positive oscillator-shaped quantum well problems in terms of the Laguerre and Jacobi polynomials are presented in this section. Detailed discussions and conclusions are given in the final section. We discuss the special cases as well as present some plots of the energy spectrum and probability distributions corresponding to certain discrete energy levels of the models under construction.

\section{Jacobi polynomials, generalized orthogonality relation on the interval $\left( a, b \right)$ and their slightly modified direct limit to the Laguerre polynomials}

Jacobi polynomials are one of the well-known and thoroughly studied continuous classical polynomials of $_2F_1$ hypergeometric type belonging to the Askey scheme of the orthogonal polynomials. They appear as a result of the positive-definite exact solution of the following second-order differential equation:

\be
\label{jp-eq1}
{\scriptstyle \left( {z^2  - 1} \right)\frac{{d^2 }}{{dz^2 }}P_n^{\left( {\alpha ,\beta } \right)} \left( \pm z \right) + \left[ \left( {\alpha  + \beta  + 2} \right)z  \pm \left( {\alpha  - \beta} \right) \right]\frac{d}{{dz}}P_n^{\left( {\alpha ,\beta } \right)} \left( \pm z \right) = n\left( {n + \alpha  + \beta  + 1} \right)P_n^{\left( {\alpha ,\beta } \right)} \left( \pm z \right),\quad n = 0,1,2, \ldots ,}
\eee
where, the Jacobi polynomials $P_n^{\left( {\alpha ,\beta } \right)} \left( z \right)$ are defined through the $_2F_1$ hypergeometric function by the following manner:

\be
\label{jp}
P_n^{\left( {\alpha ,\beta } \right)} \left( z \right) = \frac{{\left( {\alpha  + 1} \right)_n }}{{n!}}{\kern 1pt} _2 F_1 \left( {\begin{array}{*{20}c}
   { - n,n + \alpha  + \beta  + 1}  \\
   {\alpha  + 1}  \\
\end{array};\frac{{1 - z}}{2}} \right),
\eee
or it can be written down as a finite sum

\be
\label{jp2}
P_n^{\left( {\alpha ,\beta } \right)} \left( z \right) = \frac{{\left( {\alpha  + 1} \right)_n }}{{n!}}\sum\limits_{k = 0}^n {\frac{{\left( { - n} \right)_k \left( {n + \alpha  + \beta  + 1} \right)_k }}{{\left( {\alpha  + 1} \right)_k 2^k k!}}\left( {1 - z} \right)^k } .
\eee

Under condition $\alpha>-1$ and $\beta>-1$ as well as $-1<z<1$ the following continuous orthogonality relation in the finite interval holds for them:

\be
\label{orj-1}
\scriptstyle\int\limits_{ - 1}^1 {\left( {1 \mp z} \right)^\alpha  \left( {1 \pm z} \right)^\beta  P_m^{\left( {\alpha ,\beta } \right)} \left( \pm z \right)P_n^{\left( {\alpha ,\beta } \right)} \left( \pm z \right)dz}  = \frac{{2^{\alpha  + \beta  + 1} }}{{2n + \alpha  + \beta  + 1}}\frac{{\Gamma \left( {n + \alpha  + 1} \right)\Gamma \left( {n + \beta  + 1} \right)}}{{\Gamma \left( {n + \alpha  + \beta  + 1} \right)n!}}\delta _{mn} .
\eee

Let's introduce two real parameters $a$ and $b$ in such a manner that the condition $a<b$ would be satisfied. Further, via the introduction of a new variable $x$ of the following behavior:

\be
\label{x-z}
x = \frac{{a + b}}{2} \pm \frac{{a - b}}{2}z,
\eee
one obtains that eq.(\ref{jp-eq1}) will be slightly changed as follows:

\begin{equation}
\label{jp-eq2}
\begin{aligned}
\scriptstyle\left( {x - a} \right)\left( {x - b} \right)\frac{{d^2 }}{{dx^2 }}P_n^{\left( {\alpha ,\beta } \right)} \left( {\frac{{a + b - 2x}}{{b - a}}} \right) + \left[ {\left( {\alpha  + 1} \right)\left( {x - b} \right) + \left( {\beta  + 1} \right)\left( {x - a} \right)} \right]\frac{d}{{dx}}P_n^{\left( {\alpha ,\beta } \right)} \left( {\frac{{a + b - 2x}}{{b - a}}} \right) = \\ 
\scriptstyle =  n\left( {n + \alpha  + \beta  + 1} \right)P_n^{\left( {\alpha ,\beta } \right)} \left( {\frac{{a + b - 2x}}{{b - a}}} \right).
\end{aligned}
\end{equation}

One observes that due to the quasi-arbitrary nature of the parameters $\alpha$ and $\beta$, the condition $a<b$ will not be violated through the introduction of new variable $x$ via both manners shown in (\ref{x-z}). Therefore, taking into account that the following symmetry relation holds for the Jacobi polynomials: 

\be
\label{x-z1}
P_n^{\left( {\alpha ,\beta } \right)} \left( {\frac{{a + b - 2x}}{{b - a}}} \right) = \left( { - 1} \right)^n P_n^{\left( {\beta ,\alpha } \right)} \left( {\frac{{2x - a - b}}{{b - a}}} \right)
\eee
then, eq.(\ref{jp-eq2}) can also be written as follows:

\begin{equation} \label{jp-eq3}
\begin{aligned}
\scriptstyle\left( {x - a} \right)\left( {x - b} \right)\frac{{d^2 }}{{dx^2 }}P_n^{\left( {\beta ,\alpha } \right)} \left( {\frac{{2x - a - b}}{{b - a}}} \right) + \left[ {\left( {\alpha  + 1} \right)\left( {x - b} \right) + \left( {\beta  + 1} \right)\left( {x - a} \right)} \right]\frac{d}{{dx}}P_n^{\left( {\beta ,\alpha } \right)} \left( {\frac{{2x - a - b}}{{b - a}}} \right) = \\ 
\scriptstyle =  n\left( {n + \alpha  + \beta  + 1} \right)P_n^{\left( {\beta ,\alpha } \right)} \left( {\frac{{2x - a - b}}{{b - a}}} \right).
\end{aligned}
\end{equation}

Orthogonality relation (\ref{orj-1}) also changes and now holds for Jacobi polynomials $P_n^{\left( {\beta ,\alpha } \right)} \left( {\frac{{2x - a - b}}{{a - b}}} \right)$ within the interval $\left(a,b\right)$, i.e.

\be
\label{orj-2}
{\scriptstyle \int\limits_a^b {\left( {x - a} \right)^\alpha  \left( {b - x} \right)^\beta  P_m^{\left( {\beta ,\alpha } \right)} \left( {\frac{{2x - a - b}}{{b - a}}} \right)P_n^{\left( {\beta ,\alpha } \right)} \left( {\frac{{2x - a - b}}{{b - a}}} \right)dx}  = \frac{{\left( {b - a} \right)^{\alpha  + \beta  + 1} }}{{2n + \alpha  + \beta  + 1}}\frac{{\Gamma \left( {n + \alpha  + 1} \right)\Gamma \left( {n + \beta  + 1} \right)}}{{\Gamma \left( {n + \alpha  + \beta  + 1} \right)n!}}\delta _{mn} ,\quad a < x < b.}
\eee

The above-described generalized case has been discussed thoroughly in~\cite{koekoek2010}. We are going to apply such a generalization to our computations in the next section. \cite{koekoek2010} also presents the direct limit relation between the Jacobi polynomials defined through~(\ref{jp}) or (\ref{jp2}) and the Laguerre polynomials defined as follows:

\be
\label{lp}
L_n^{\left( \alpha  \right)} \left( z \right) = \frac{{\left( {\alpha  + 1} \right)_n }}{{n!}}\,_1 F_1 \left( {\begin{array}{*{20}c}
   { - n}  \\
   {\alpha  + 1}  \\
\end{array};z} \right) = \frac{1}{{n!}}\sum\limits_{k = 0}^n {\frac{{\left( { - n} \right)_k }}{{k!}}\left( {\alpha  + k + 1} \right)_{n - k} z^k } .
\eee

Then, if we take $z=1-2 \beta^{-1}x$ in the definition of the Jacobi polynomials~(\ref{jp}) and let $\beta \to \infty$, one obtains that

\be
\label{jp-lp-1}
\mathop {\lim }\limits_{\beta  \to \infty } P_n^{\left( {\alpha ,\beta } \right)} \left( {1 - \frac{{2x}}{\beta }} \right) = L_n^{\left( \alpha  \right)} \left( x \right).
\eee

We are going to change slightly this known limit relation between the Jacobi and Laguerre polynomials. First of all, we connect $\alpha$ and $\beta$ parameters of the Jacobi polynomials as follows:

\be
\label{ba}
\beta  = \frac{b}{a}\alpha .
\eee

Next, letting $b \to \infty$ one obtains that the following limit relation between the Jacobi and Laguerre polynomials holds:

\be
\label{jp-lp-2}
\mathop {\lim }\limits_{b \to \infty } P_n^{\left( {\alpha ,\frac{b}{a}\alpha } \right)} \left( {1-2\frac{{x-a}}{{b - a}}} \right) = L_n^{\left( \alpha  \right)} \left( {\frac{\alpha }{a}x - \alpha } \right).
\eee

One observes that limit relation~(\ref{jp-lp-2}) is a slightly modified version of~(\ref{jp-lp-1}) and it is valid for the Jacobi polynomials being appeared in the differential equation~(\ref{jp-eq2}). Its application to this equation yields the following differential equation for the Laguerre polynomials $L_n^{\left( \alpha  \right)} \left( {\frac{\alpha }{a}x - \alpha } \right)$:

\be
\label{lp-eq-1}
\left( {x - a} \right)\frac{{d^2 }}{{dx^2 }}L_n^{\left( \alpha  \right)} \left( {\frac{\alpha }{a}x - \alpha } \right) + \left( {2\alpha  + 1 - \frac{\alpha }{a}x} \right)\frac{d}{{dx}}L_n^{\left( \alpha  \right)} \left( {\frac{\alpha }{a}x - \alpha } \right) =  - \frac{\alpha }{a}nL_n^{\left( \alpha  \right)} \left( {\frac{\alpha }{a}x - \alpha } \right).
\eee

We are going to apply such a generalization to our computations in the next section, too. This solution in terms of the Laguerre polynomials will be directly related to our oscillator-shaped quantum well model being confined only from the right side with the infinitely high wall.

\section{Position-dependent mass Schr\"odinger equation with the positive oscillator-shaped quantum well potential}

Let's consider a quantum system confined only from the right side at a position positive value $x=a$ with an infinitely high wall. In fact, this wall forms a model of the semi-infinitely deep quantum well of the arbitrary behavior of one of its walls. Attractivity of the problem under consideration increases enormously if one manages to succeed with an analytical solution to its Schr\"odinger equation under a certain type of potential energies. We are going to demonstrate how one can obtain an exact solution to the Schr\"odinger equation with the completely positive oscillator-shaped quantum well potential. The notion of 'the completely positive oscillator-shaped quantum well potential' means that the non-relativistic quantum system described by the Schr\"odinger equation is valid only for certain positive values of the position.

Our starting point is the Schr\"odinger equation in one dimension:

\be
\label{se-1}
\left[ {\frac{{\hat p_x }}{{2m_0 }} + V\left( x \right)} \right]\psi \left( x \right) = E\psi \left( x \right).
\eee

Here, $\psi \left( x \right)$ is the wavefunction of the stationary states of the quantum system under study, $m_0$ is the constant mass and $E$ is its energy. One-dimensional momentum operator $\hat p_x$ can be defined within the canonical or non-canonical approach. We are going to deal with the canonical approach, within which the momentum operator is defined as

\[
\hat p_x  =  - i\hbar \frac{d}{{dx}}.
\]

Its substitution at (\ref{se-1}) leads to the following second-order differential equation:

\be
\label{se-2}
\left[ { - \frac{{\hbar ^2 }}{{2m_0 }}\frac{{d^2 }}{{dx^2 }} + V\left( x \right)} \right]\psi \left( x \right) = E\psi \left( x \right).
\eee

Now, one needs to introduce the analytical expression of the positive oscillator-shaped quantum well potential $V\left( x \right)$. One of the ways is a direct substitution of the well-known harmonic oscillator potential with the constant mass $m_0$ and angular frequency $\omega$ at eq.(\ref{se-2}) and then, to solve corresponding second-order equation within the finite positive values of the position $a<x<+\infty$ with additional confinement condition of existence of an infinitely high wall at position $x=a$. It is well known that in this case eq.(\ref{se-2}) will not be solved exactly. Therefore, we are going with a different way through the replacement of the constant mass with a position-dependent one: $m_0 \to M\left( x \right)$. The mass varying with position is a well-known conception in physics. Its origin can be traced back to seminal electron tunneling experiments~\cite{giaever1960a,giaever1960b} and their theoretical explanation within the approach of the band structure varying with position~\cite{harrison1961}. Moreover, the approach was developed to the assumption that varying band structure of the solid compounds can lead also to the effective mass varying with position and the following replacement for the non-relativistic kinetic energy operator was proposed~\cite{bendaniel1966}:

\[
 - \frac{{\hbar ^2 }}{{2m_0 }}\frac{{d^2 }}{{dx^2 }} \to  - \frac{{\hbar ^2 }}{2}\frac{d}{{dx}}\frac{1}{{M\left( x \right)}}\frac{d}{{dx}}.
\]

Such a replacement also preserves the Hermiticity property of the kinetic energy operator. Its substitution at (\ref{se-2}) yields:

\be
\label{se-3}
\left[ { - \frac{{\hbar ^2 }}{2}\frac{d}{{dx}}\frac{1}{{M\left( x \right)}}\frac{d}{{dx}} + V\left( x \right)} \right]\psi \left( x \right) = E\psi \left( x \right).
\eee

Recently, it was shown that this equation can be considered as a powerful tool for simultaneous study of the movement of an electron in a semiconductor quantum well and the propagation of electromagnetic waves in a dielectric guide~\cite{barsan2022}.

Taking into account that

\[
\frac{d}{{dx}}\frac{1}{M}\frac{d}{{dx}} = \frac{1}{M}\frac{{d^2 }}{{dx^2 }} - \frac{{M'}}{{M^2 }}\frac{d}{{dx}},\quad M \equiv M\left( x \right),
\]
one can write down eq.(\ref{se-3}) as follows:

\be
\label{se-4}
\left[ { - \frac{{\hbar ^2 }}{{2M}}\frac{{d^2 }}{{dx^2 }} + \frac{{\hbar ^2 }}{2}\frac{{M'}}{{M^2 }}\frac{d}{{dx}} + V\left( x \right)} \right]\psi \left( x \right) = E\psi \left( x \right).
\eee

This second-order differential equation can be solved exactly if one defines analytical expressions of the potential $V\left( x \right)$ and position-dependent mass $M\left( x \right)$. As we already highlighted that our potential is almost the same as one-dimensional harmonic oscillator potential, but with the mass varying with the position. Analytical expression of the mass should be of such form that it will go to positive infinity at values of position $a$, creating an infinitely high impenetrable wall as well as preserving completely positive harmonic behavior of the oscillator for position values $x>a$. The main feature of this oscillator is that it never approaches its equilibrium state being completely positive. Of course, another important condition for its introduction is the exact solubility of eq.(\ref{se-4}).

Briefly, the following analytical expression of the position-dependent mass $M\left( x \right)$

\[
M\left( x \right) = \frac{{am_0 }}{{ {x - a} }},
\]
and potential energy $V\left( x \right)$

\[
V\left( x \right) = \frac{{M\omega^2 x^2 }}{2},\quad a < x,
\]
is proposed.

Their substitution at (\ref{se-4}) and further simple mathematical tricks yields:

\be
\label{se-5}
\psi '' + \frac{1}{{x - a}}\psi ' - \frac{{\lambda _0 ^4 a^2 x^2  - c_0 x + ac_0 }}{{\left( {x - a} \right)^2 }}\psi  = 0,
\eee
where the notations $\frac{{d\psi }}{{dx}} \equiv \psi '$ and $\frac{{d^2 \psi }}{{dx^2 }} \equiv \psi ''$ are introduced for simplicity.

Here, also

\begin{equation} \label{c0-lambda}
 c_0  =  \frac{{2m_0 aE}}{{\hbar ^2 }}, \qquad \lambda _0 = \sqrt{\frac{{m_0 \omega }}{{\hbar }}},
\end{equation}
are introduced.

We look for exact solution of eq.(\ref{se-5}) in the form of

\be
\label{wf-1}
\psi \left( x \right) = \varphi \left( x \right)y\left( x \right),
\eee
where $y \left( x \right)$ is the hypergeometric function part of the wavefunction that defines its positive oscillator behavior at position values $x>a$ as well as $\varphi \left( x \right)$ is its weight function that is going to suppress the wavefunction to vanish under the boundary condition $x=a$. Substitution of (\ref{wf-1}) at eq.(\ref{se-5}) yields:

\be
\label{se-6}
\varphi y'' + \left[ {2\varphi ' + \frac{1}{{x - a}}\varphi } \right]y' + \left[ {\varphi '' + \frac{1}{{x - a}}\varphi ' - \frac{{\lambda _0 ^4 a^2 x^2  - c_0 x + ac_0 }}{{\left( {x - a} \right)^2 }}\varphi } \right]y = 0.
\eee

Taking into account vanishing behavior of the wavefunction at values $x=a$ and $x \to +\infty$, one introduces $\varphi \left( x \right)$ of the following semi-general form:

\be
\label{phi-1}
\varphi \left( x \right) = \left( {x - a} \right)^A e^{Bx} .
\eee

Here, $A$ and $B$ are arbitrary parameters, which values will be defined below being related to the general behavior of the wavefunction both between walls and at borders.

Simple computations allow to obtain that,

\be
\label{phid-1}
\varphi ' = \frac{{Bx + A - aB}}{{x - a}}\varphi 
\eee
and

\be
\label{phid-2}
\varphi '' = \frac{{B^2 x^2  + 2B\left( {A - aB} \right)x + \left( {A - aB} \right)^2  - A}}{{\left( {x - a} \right)^2 }}\varphi .
\eee

One observes from (\ref{phid-2}) that the positive oscillator behavior of the wavefunction between two walls or polynomial solution of eq.(\ref{se-6}) is possible only if the following system of equations hold:

\[
\left\{ \begin{array}{l}
 B^2  - \lambda _0 ^4 a^2  = 0 , \\ 
 c_0  + 2B\left( {A - aB} \right) + B = \mu , \\ 
 c_0  + B - \frac{{\left( {A - aB} \right)^2 }}{a} = \mu . \\ 
 \end{array} \right.
\]

Here, $\mu$ is an arbitrary real parameter. The above system of equations can be easily solved leading to the following values for $A$, $B$, and $\mu$:

\[
A = \varepsilon _1 \lambda _0 ^2 a^2,\; B = \varepsilon _2 \lambda _0 ^2 a,\; \mu  = c_0  + 2\varepsilon _2 \lambda _0 ^2 a\left( {\varepsilon _1  - \varepsilon _2 } \right)\lambda _0 ^2 a^2  + \varepsilon _2 \lambda _0 ^2 a ,\quad \varepsilon _{1,2} =  \pm 1.
\]

Then, one observes for (\ref{phi-1}) that

\be
\label{phi-2}
\varphi \left( x \right) = \left( {x - a} \right)^{\varepsilon _1 \lambda _0 ^2 a^2 } e^{\varepsilon _2 \lambda _0 ^2 ax} ,
\eee
and as a consequence of it eq.(\ref{se-6}) reduces to the equation for $y\left( x \right)$ of the following form:

\be \label{se-7}
\scriptstyle\left( {x - a} \right)y'' + \left[ {2\left( {\varepsilon _1  - \varepsilon _2 } \right)\lambda _0 ^2 a^2  + 1 + 2\varepsilon _2 \lambda _0 ^2 ax} \right]y' + \left[ {c_0  + 2\varepsilon _2 \lambda _0 ^2 a\left( {\varepsilon _1  - \varepsilon _2 } \right)\lambda _0 ^2 a^2  + \varepsilon _2 \lambda _0 ^2 a} \right]y = 0.
\eee

Taking into account that parameters $\varepsilon_{1,2}$ can be both positive and negative, then the four possible scenarios can be constructed. However, only the vanishing behavior of the wavefunction with one of these values will be satisfied. This values are $\varepsilon_1=+1$ and $\varepsilon_2=-1$. One simplifies for (\ref{phi-2}) as follows:

\be
\label{phi-3}
\varphi \left( x \right) = \left( {x - a} \right)^{\lambda _0 ^2 a^2 } e^{ - \lambda _0 ^2 ax},
\eee
as well as eq.(\ref{se-7}) also slightly simplifies to the following form:

\be \label{se-8}
\left( {x - a} \right)y'' + \left( {4\lambda _0 ^2 a^2  + 1 - 2\lambda _0 ^2 ax} \right)y' + \left( {c_0  - 4\lambda _0 ^4 a^3  - \lambda _0 ^2 a} \right)y = 0.
\eee

This second-order differential equation for $y \left( x \right)$ is already known to us from the previous section. One observes its similarity with eq.(\ref{lp-eq-1}). Then, the comparison of both (\ref{se-8}) and (\ref{lp-eq-1}) yields the following analytical expression of the energy spectrum

\be
\label{en1}
E_n  =  \hbar \omega \left( {n + \frac{1}{2}} \right) + 2m_0 \omega ^2a^2,
\eee
and orthonormalized wavefunctions in terms of the Laguerre polynomials as follows:

\be
\label{wf-lp}
\psi _n \left( x \right) = C_n \left( {x - a} \right)^{\lambda _0 ^2 a^2 } e^{ - \lambda _0 ^2 ax} L_n^{\left( {2\lambda _0 ^2 a^2 } \right)} \left( {2\lambda _0 ^2 a\left( {x - a} \right)} \right),
\eee
where the orthonormalized coefficient of the following analytical expression 

\be
\label{nf-l}
C_n  = \left( -1 \right)^n e^{\lambda _0 ^2 a^2 } \left( {2\lambda _0 ^2 a } \right)^{\lambda _0 ^2 a^2  + \frac{1}{2}} \sqrt {\frac{{n!}}{{\Gamma \left( {n + 2\lambda _0 ^2 a^2  + 1} \right)}}} 
\eee
can be easily deduced from the following orthogonality relation for the Laguerre polynomials:

\[
\scriptstyle\int\limits_a^\infty  {e^{ - 2\lambda _0 ^2 ax} \left( {x - a} \right)^{2\lambda _0 ^2 a^2 } L_m^{\left( {2\lambda _0 ^2 a^2 } \right)} \left( {2\lambda _0 ^2 a\left( {x - a} \right)} \right)L_n^{\left( {2\lambda _0 ^2 a^2 } \right)} \left( {2\lambda _0 ^2 a\left( {x - a} \right)} \right)dx}  = \frac{{\Gamma \left( {n + 2\lambda _0 ^2 a^2  + 1} \right)}}{{e^{2\lambda _0 ^2 a^2 } \left( {2\lambda _0 ^2 a } \right)^{2\lambda _0 ^2 a^2  + 1} n!}}\delta _{mn} .
\]

Next, let's consider a more general quantum system, i.e. the system confined to the cavity within the two positively infinite high walls. These walls are located in positive position values $x=a$ and $x=b$ with initial condition $a<b$. One deduces that these two walls are separated by a distance $b-a$ and in fact, they form a model of the infinitely deep quantum well of the arbitrary behavior. In other words, such behavior can have an arbitrary profile depending on the analytical expression of the potential energy being applied between two walls. We are going to demonstrate how one can obtain an exact solution to the Schr\"odinger equation with the completely positive oscillator-shaped quantum well potential for the case, if the quantum system under construction is confined within the two infinitely high walls of positive values and behaves itself as an oscillator between these walls.

In this general case, the following analytical expression of the position-dependent mass $M\left( x \right)$

\[
M\left( x \right) = \frac{{abm_0 }}{{ \left({x - a}\right)\left({b - x}\right) }},
\]
and potential energy $V\left( x \right)$

\[
V\left( x \right) = \frac{{M\omega^2 x^2 }}{2},\quad a < x < b,
\]
is proposed.

Again, their substitution at (\ref{se-4}) and further simple mathematical tricks yields:

\be
\label{se2-5}
\psi '' - \frac{{2x - \left( {a + b} \right)}}{{\left( {x - a} \right)\left( {b - x} \right)}}\psi ' - \frac{{c_2 x^2  - \left( {a + b} \right)c_1 x + abc_1 }}{{\left( {x - a} \right)^2 \left( {b - x} \right)^2 }}\psi  = 0,
\eee
where, 

\begin{equation} \label{c1-c2}
\begin{aligned}
 c_1  =&  bc_0 = \frac{{2m_0 abE}}{{\hbar ^2 }}, \\ 
 c_2  =& c_1  + \lambda _0 ^4 a^2 b^2 ,
\end{aligned}
\end{equation}
are introduced for simplicity.

We look for exact solution of eq.(\ref{se2-5}) in the form of

\be
\label{wf2-1}
\psi \left( x \right) = \varphi \left( x \right)y\left( x \right),
\eee
where $y \left( x \right)$ is the hypergeometric function part of the wavefunction that defines its positive oscillator behavior between two walls and $\varphi \left( x \right)$ is its weight function that is going to suppress the wavefunction to vanish under the boundary conditions $x=a$ and $x=b$. Substitution of (\ref{wf2-1}) at eq.(\ref{se2-5}) yields:

\be
\label{se2-6}
\scriptstyle\varphi y'' + \left[ {2\varphi ' - \frac{{2x - \left( {a + b} \right)}}{{\left( {x - a} \right)\left( {b - x} \right)}}\varphi } \right]y' + \left[ {\varphi '' - \frac{{2x - \left( {a + b} \right)}}{{\left( {x - a} \right)\left( {b - x} \right)}}\varphi ' - \frac{{c_2 x^2  - \left( {a + b} \right)c_1 x + abc_1 }}{{\left( {x - a} \right)^2 \left( {b - x} \right)^2 }}\varphi } \right]y = 0.
\eee

Taking into account vanishing behavior of the wavefunction at values $x=a$ and $x=b$, one introduces $\varphi \left( x \right)$ of the following semi-general form:

\be
\label{phi2-1}
\varphi \left( x \right) = \left( {x - a} \right)^A \left( {b - x} \right)^B .
\eee

Here, $A$ and $B$ again are arbitrary parameters, which values will be defined below being related to the general behavior of the wavefunction both between walls and at borders.

Simple computations allow to obtain that,

\be
\label{phid2-1}
\varphi ' = \frac{{\left( {bA + aB} \right) - \left( {A + B} \right)x}}{{\left( {x - a} \right)\left( {b - x} \right)}}\varphi 
\eee
and

\be
\label{phid2-2}
\scriptstyle\varphi '' = \frac{{\left[ {\left( {bA + aB} \right) - \left( {A + B} \right)x} \right]^2  + \left[ {\left( {bA + aB} \right) - \left( {A + B} \right)x} \right]\left[ {2x - \left( {a + b} \right)} \right] - \left( {A + B} \right)\left( {x - a} \right)\left( {b - x} \right)}}{{\left( {x - a} \right)^2 \left( {b - x} \right)^2 }}\varphi .
\eee

One observes from (\ref{phid2-2}) that the positive oscillator behavior of the wavefunction between two walls or polynomial solution of eq.(\ref{se2-6}) is possible only if the following system of equations hold:

\[
\left\{ \begin{array}{l}
 c_2  - \left( {A + B} \right)^2  - \left( {A + B} \right)  = \mu , \\ 
 c_1  - \left( {A + B} \right) - \frac{{2\left( {bA + aB} \right)\left( {A + B} \right)}}{{a + b}}  = \mu , \\ 
 c_1  - \left( {A + B} \right) - \frac{{\left( {bA + aB} \right)^2 }}{{ab}}  = \mu . \\ 
 \end{array} \right.
\]

Here, $\mu$ again is an arbitrary real parameter. The above system of equations again can be easily solved leading to the following values for $A$, $B$ and $\mu$:

\[
\scriptstyle A = \varepsilon _1 \lambda _0 ^2 \frac{{a^2 b}}{{a - b}},\; B = \varepsilon _2 \lambda _0 ^2 \frac{{ab^2 }}{{a - b}},\; \mu  = c_1  - \lambda _0 ^2 \frac{{ab}}{{a - b}}\left( {\varepsilon _1 a + \varepsilon _2 b} \right) - 2\lambda _0 ^4 \frac{{a^3 b^3 }}{{\left( {a - b} \right)^2 }}\left( {1 + \varepsilon _1 \varepsilon _2 } \right) ,\quad \varepsilon _{1,2} =  \pm 1.
\]

Then, one observes for (\ref{phi2-1}) that

\be
\label{phi2-2}
\varphi \left( x \right) = \left( {x - a} \right)^{\varepsilon _1 \lambda _0 ^2 \frac{{a^2 b}}{{a - b}}} \left( {b - x} \right)^{\varepsilon _2 \lambda _0 ^2 \frac{{ab^2 }}{{a - b}}} ,
\eee
and as a consequence of it eq.(\ref{se2-6}) reduces to the equation for $y\left( x \right)$ of the following form:

\be \label{se2-7}
\begin{aligned}
 \left( {x - a} \right)\left( {b - x} \right)y'' +& \left[ {2\lambda _0 ^2 \frac{{a^2 b^2 }}{{a - b}}\left( {\varepsilon _1  + \varepsilon _2 } \right) + \left( {a + b} \right) - 2\left( {\varepsilon _1 \lambda _0 ^2 \frac{{a^2 b}}{{a - b}} + \varepsilon _2 \lambda _0 ^2 \frac{{ab^2 }}{{a - b}} + 1} \right)x} \right]y' \\ 
  +& \left[ {c_1  - \lambda _0 ^2 \frac{{ab}}{{a - b}}\left( {\varepsilon _1 a + \varepsilon _2 b} \right) - 2\lambda _0 ^4 \frac{{a^3 b^3 }}{{\left( {a - b} \right)^2 }}\left( {1 + \varepsilon _1 \varepsilon _2 } \right)} \right]y = 0.
\end{aligned}
\eee

Now, still the parameters $\varepsilon_{1,2}$ can be both positive and negative, hence the four possible scenarios can be constructed. However, it is clear that only the vanishing behavior of the wavefunction with one of these values will be satisfied. This value is $\varepsilon_{1,2}=-1$. One simplifies for (\ref{phi2-2}) as follows:

\be
\label{phi2-3}
\varphi \left( x \right) = \left( {x - a} \right)^{\lambda _0 ^2 \frac{{a^2 b}}{{b - a}}} \left( {b - x} \right)^{\lambda _0 ^2 \frac{{ab^2 }}{{b - a}}},
\eee
as well as eq.(\ref{se2-7}) also slightly will be simplified to the following form:

\be \label{se2-8}
\scriptstyle \left( {x - a} \right)\left( {b - x} \right)y'' + \left[ {2\left( {\lambda _0 ^2 ab\frac{{a + b}}{{a - b}} - 1} \right)x - 4\lambda _0 ^2 \frac{{a^2 b^2 }}{{a - b}} + \left( {a + b} \right)} \right]y' + \left[ {c_1  + \lambda _0 ^2 ab\frac{{a + b}}{{a - b}} - 4\lambda _0 ^4 \frac{{a^3 b^3 }}{{\left( {a - b} \right)^2 }}} \right]y = 0.
\eee

This second-order differential equation for $y \left( x \right)$ is also already known to us from the previous section. One observes its similarity with eq.(\ref{jp-eq3}). Then, the comparison of both (\ref{se2-8}) and (\ref{jp-eq3}) yields the following analytical expression of the energy spectrum

\be
\label{en21}
E_n  =  \frac{{b + a}}{{b - a}}\hbar \omega \left( {n + \frac{1}{2}} \right) + \frac{{\hbar ^2 }}{{2m_0 ab}}n\left( {n + 1} \right) + 2m_0 \omega ^2 \frac{{a^2 b^2 }}{{\left( {b - a} \right)^2 }},
\eee
and orthonormalized wavefunctions in terms of the Jacobi polynomials as follows:

\be
\label{wf-jp}
\psi _n \left( x \right) = C_n \left( {x - a} \right)^{\lambda _0 ^2 \frac{{a^2 b}}{{b - a}}} \left( {b - x} \right)^{\lambda _0 ^2 \frac{{ab^2 }}{{b - a}}} P_n^{\left( {2\lambda _0 ^2 \frac{{a b^2}}{{b - a}},2\lambda _0 ^2 \frac{{a^2 b}}{{b - a}}} \right)} \left( {\frac{{2x - a - b}}{{b - a}}} \right),
\eee
where the orthonormalization coefficient of the following analytical expression can be easily deduced from (\ref{orj-2}):

\be
\label{nf-j}
C_n  = \left( -1 \right)^n \left( {b - a} \right)^{ - \lambda _0 ^2 ab\frac{{b + a}}{{b - a}} - \frac{1}{2}} \sqrt {\frac{{\left( {2n + 2\lambda _0 ^2 ab\frac{{b + a}}{{b - a}} + 1} \right)\Gamma \left( {n + 2\lambda _0 ^2 ab\frac{{b + a}}{{b - a}} + 1} \right)n!}}{{\Gamma \left( {n + 2\lambda _0 ^2 \frac{{a^2 b}}{{b - a}} + 1} \right)\Gamma \left( {n + 2\lambda _0 ^2 \frac{{ab^2 }}{{b - a}} + 1} \right)}}}.
\eee

Exact analytical expressions of the wavefunctions (\ref{wf-lp}) and (\ref{wf-jp}), as well as analytical expressions of the energy spectrum (\ref{en1}) and (\ref{en21}), are the main goal that we needed to achieve. Now, we are going to discuss briefly the possible connection between these analytical expressions as well as their behavior under special conditions.

\section{Discussions and Conclusions}

We managed to solve exactly both equations (\ref{se-5}) and (\ref{se2-5}). One observes from obtained analytical expressions of the eigenvalues of these second-order differential equations, the behavior of both of them is drastically different. First of all, the energy spectrum (\ref{en1}) corresponding to the model with only one infinitely high wall is equidistant. However, the energy spectrum (\ref{en21}) corresponding to the model located within two infinitely high walls exhibits non-equidistant behavior. Also, despite that the first model is confined to the positive region of position $a<x$, its energy spectrum (\ref{en1}) is almost similar to the well-known non-relativistic quantum harmonic oscillator with the constant mass~\cite{bloch1997,landau1991,moshinsky1996}. Its first term completely overlaps with the known oscillator energy spectrum. The only difference in energy spectrum (\ref{en1}) is the existence of its additional second term that depends on the so-called confinement parameter $a$. It seems that if letting to $a \to 0$, then the non-relativistic quantum harmonic oscillator model will be directly recovered through simple mathematical computations. However, it is a misassumption. Therefore, one needs to analyze some details of the energy spectrum (\ref{en21}), too. It consists of three terms. The first term does not overlap but is similar to the known oscillator energy spectrum. The second term adds non-equidistant behavior and the third term is simply some constant that does not contribute anything to the general non-equidistant behavior of the energy spectrum (\ref{en21}). In general, all three terms depend on parameters $a$ and $b$. Then, the case $a \to 0$ violates its such a general behavior through the second term. Parameter $a$ is in its denominator. Therefore, this term goes to infinity. From a physics viewpoint, masses of both models changing by position simply become zero if $a \to 0$. Therefore, both models restricted within the positive region have a number of singularities, which never allow them to recover the well-known non-relativistic quantum harmonic oscillator model with the wavefunctions expressed through the Hermite polynomials.

\begin{figure}[h!]
\includegraphics[scale=0.30]{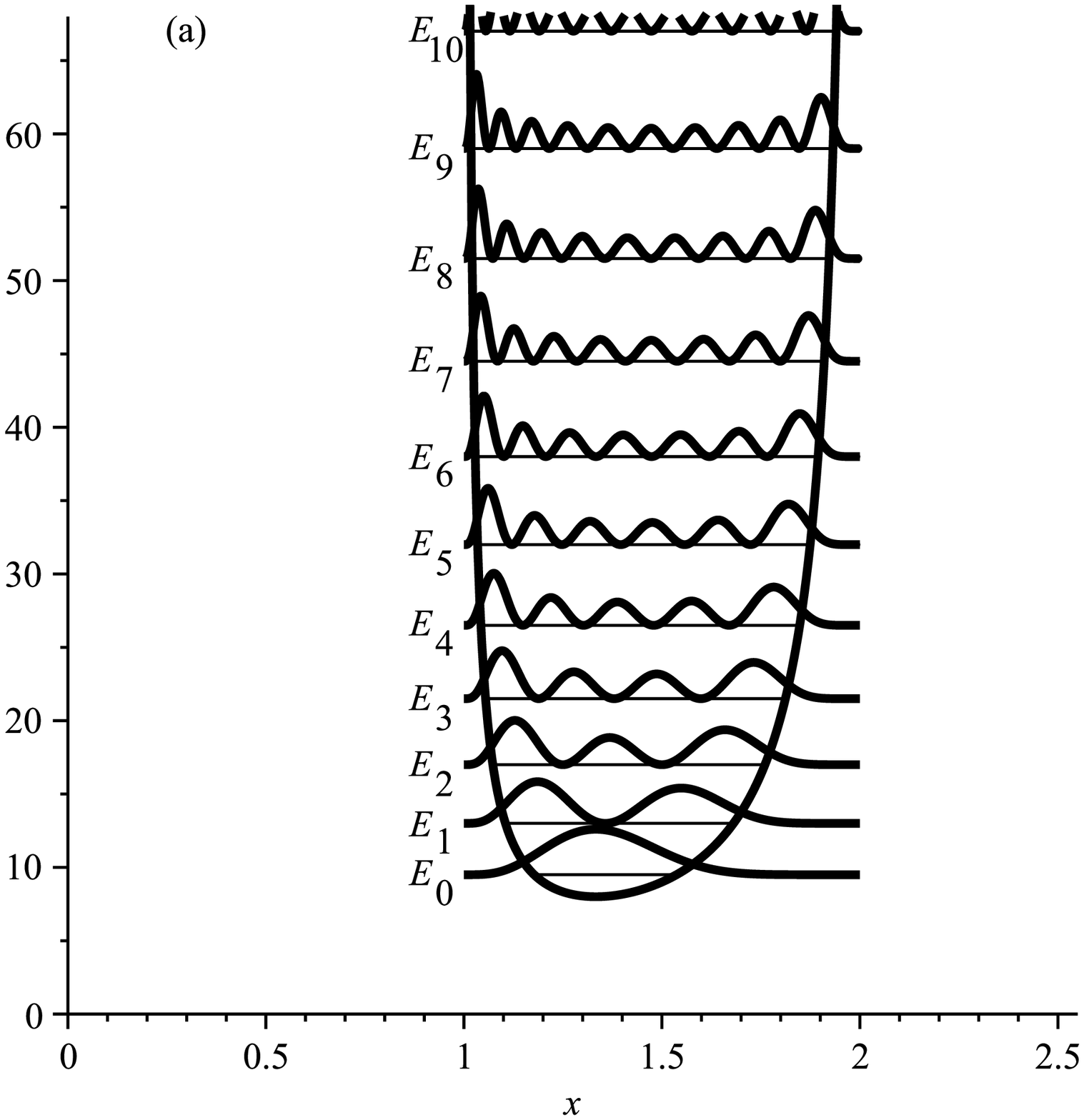}
\hspace{0.1cm} 
\includegraphics[scale=0.30]{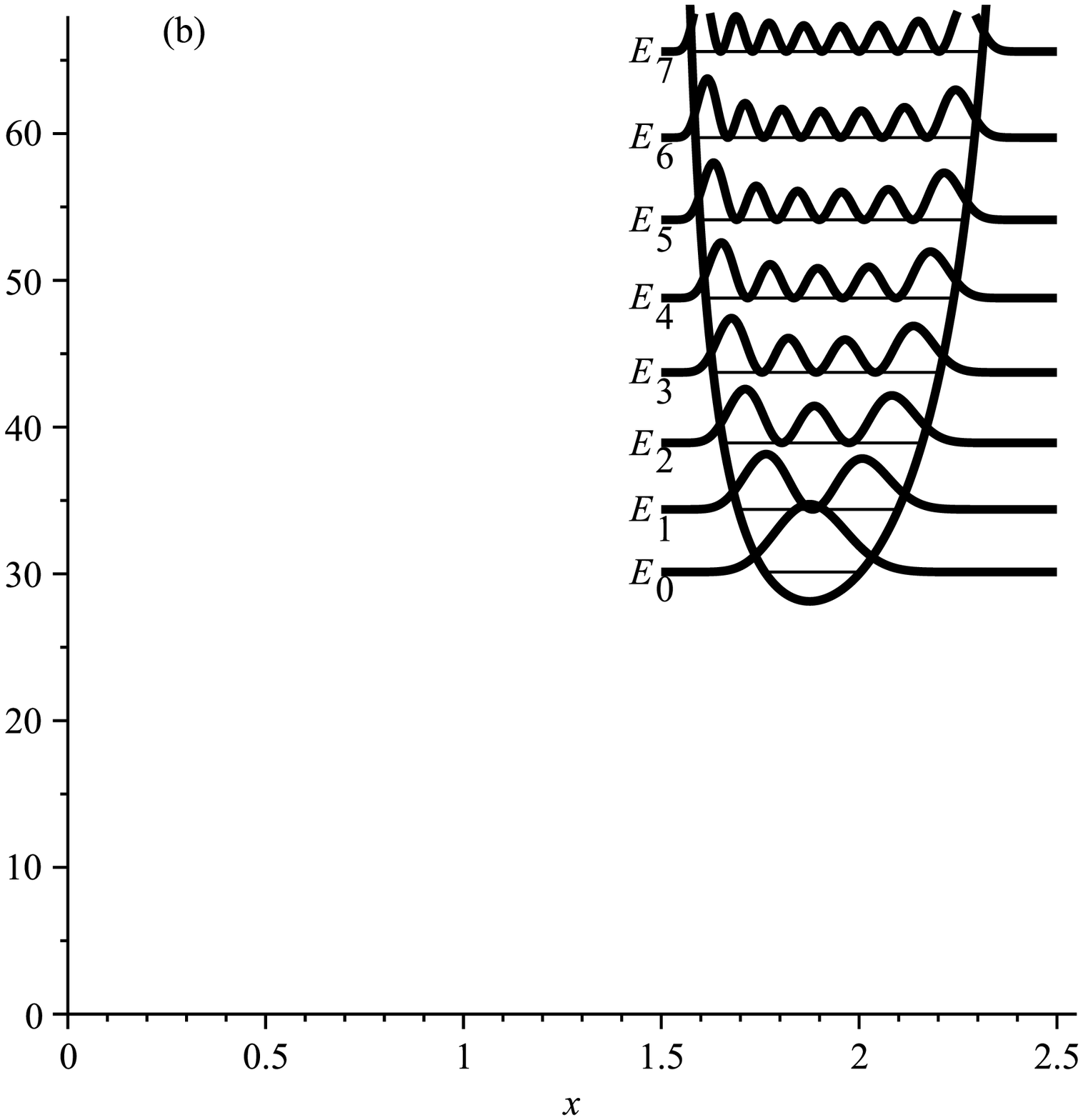}
\caption{Dependence of the probability densities $\left| \psi _n \left( x \right) \right|^2$ generated via (\ref{wf-jp}) and non-equidistant energy levels~(\ref{en21}) from the confinement parameters $a$ and $b$ for the ground and a)  10 excited states if $a=1$ and $b=2$; b)  7 excited states if $a=1.5$ and $b=2.5$  ($m_0=\omega=\hbar=1$).}
\label{fig.1}
\end{figure}

In fig.\ref{fig.1}, we depicted dependence of the probability densities $\left| \psi _n \left( x \right) \right|^2$ generated via (\ref{wf-jp}) and non-equidistant energy levels~(\ref{en21}) corresponding to them from the confinement parameters $a$ and $b$. In order to keep the overlapping scales of the plot, we visualize only a restricted number of the stationary state. There are ground plus $10$ excited states if $a=1$ and $b=2$ as well as ground plus $7$ excited states if $a=1.5$ and $b=2.5$. For simplicity, the measurement system $m_0=\omega=\hbar=1$ is chosen, too. One observes Gaussian-like behavior for the ground state probability densities $\left| \psi _0 \left( x \right) \right|^2$ for both values of $a$ and $b$. Another feature that can be observed from these plots is the exhibition by the system more probability close to $x \to a$ than $x \to b$. Of course, the main feature of the model under construction is that its equilibrium state is always different than zero. This is a consequence of a completely positive definition of the model in the initial state.

\begin{figure}[h!]
\includegraphics[scale=0.30]{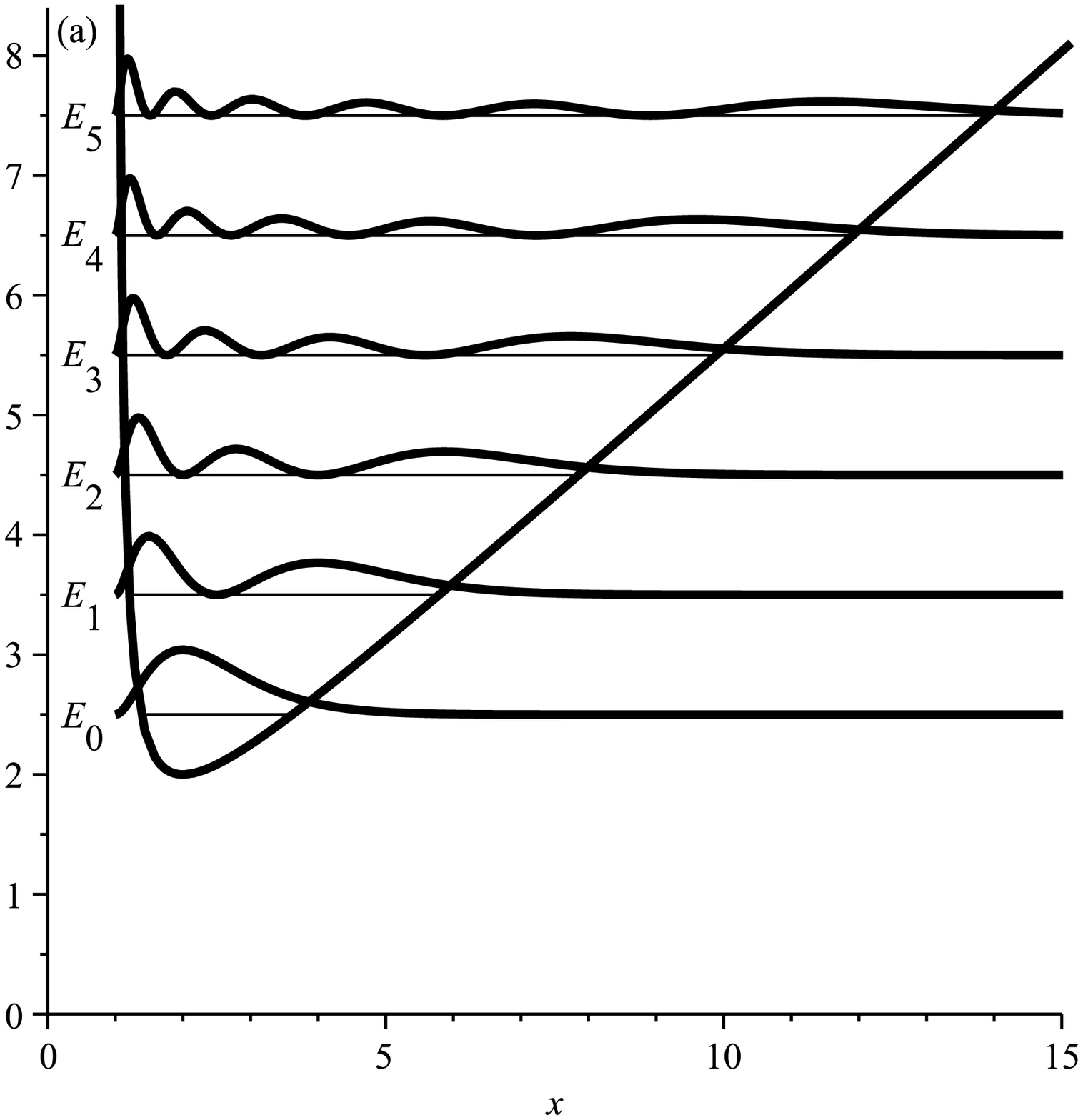}
\hspace{0.1cm} 
\includegraphics[scale=0.30]{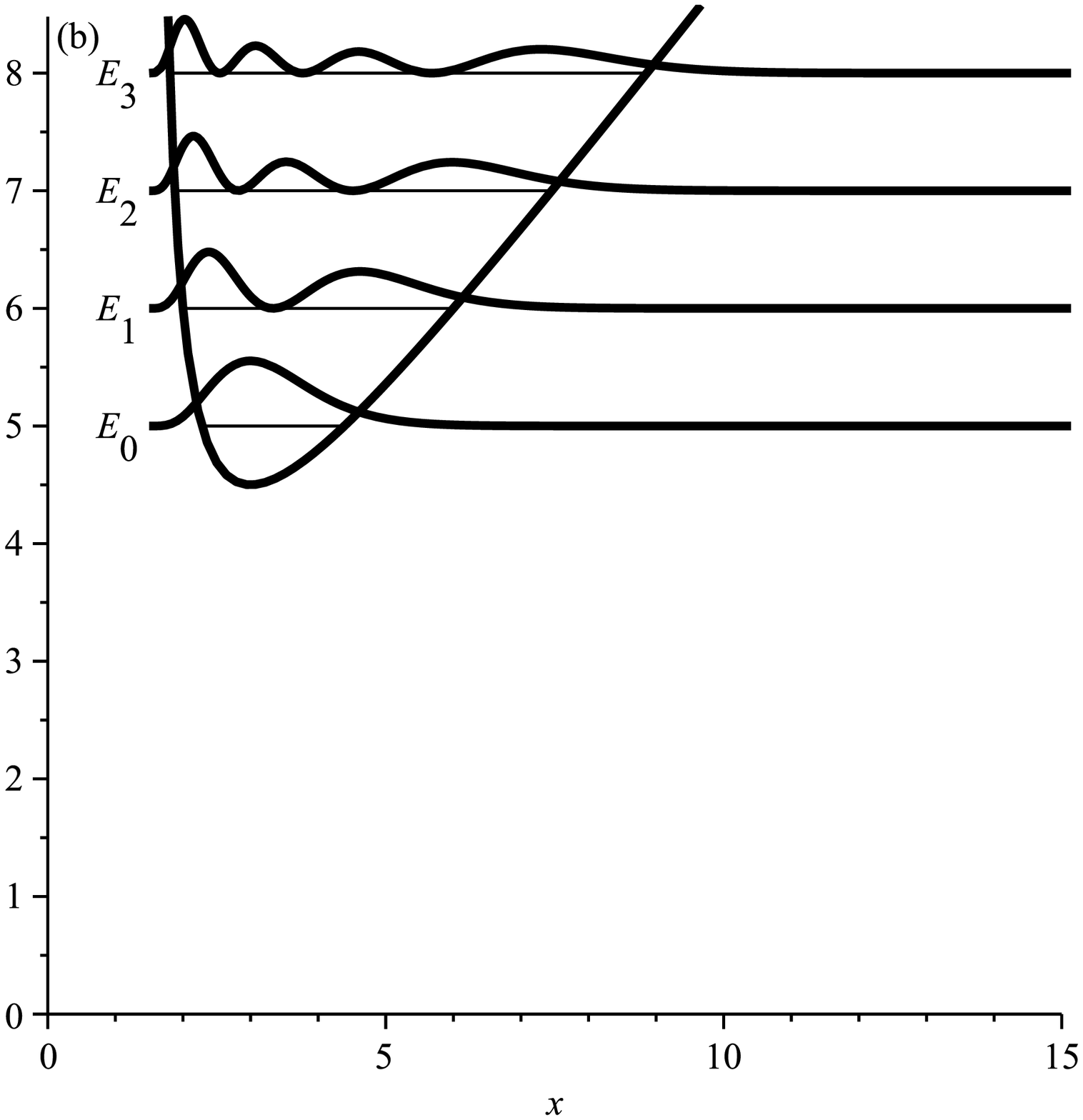}
\caption{Dependence of the probability densities $\left| \psi _n \left( x \right) \right|^2$ generated via (\ref{wf-lp}) and equidistant energy levels~(\ref{en1}) from the confinement parameter $a$ for the ground and a)  5 excited states if $a=1$; b)  3 excited states if $a=1.5$  ($m_0=\omega=\hbar=1$).}
\label{fig.2}
\end{figure}

The case, when $b \to +\infty$ is depicted in fig.\ref{fig.2}. We visualize dependence of the probability densities $\left| \psi _n \left( x \right) \right|^2$ generated via (\ref{wf-lp}) and equidistant energy levels~(\ref{en1}) corresponding to these probability densities from the confinement parameter $a$. Mathematically, the limit $b \to +\infty$ results in the disappearance of the second term in (\ref{en21}) as well as simplifications of its first and third terms. One observes equidistant energy levels as well as higher probability densities close to the value of the position $x \to a$. Also,  Gaussian-like behavior for the ground state probability densities $\left| \psi _0 \left( x \right) \right|^2$ still can be observed here.  

Recovery of the polynomial part of the wavefunction (\ref{wf-lp}) from the wavefunction (\ref{wf-jp}) under the limit $b \to +\infty$ is based on the slightly modified limit relation between Jacobi and Laguerre polynomials, introduced via (\ref{jp-lp-2}). The recovery of the weight function of the wavefunction (\ref{wf-lp}) can be performed through the following basic limit relations:

\[
\mathop {\lim }\limits_{b \to \infty } \left( {x - a} \right)^{\lambda _0 ^2 \frac{{a^2 b}}{{b - a}}}  = \left( {x - a} \right)^{\lambda _0 ^2 a^2 } ,
\]
\[
\mathop {\lim }\limits_{b \to \infty } \Gamma \left( {n + 2\lambda _0 ^2 \frac{{a^2 b}}{{b - a}} + 1} \right) = \Gamma \left( {n + 2\lambda _0 ^2 a^2  + 1} \right),
\]
and

\[
\mathop {\lim }\limits_{b \to \infty } \sqrt {\frac{{2n + 2\lambda _0 ^2 ab + 1}}{{b - a}}}  = \sqrt {2\lambda _0 ^2 a} ,
\] 
as well as by applying Stirling's approximation for the gamma function $\Gamma \left( {z + 1} \right)\approx \sqrt {2\pi } x^{x + \frac{1}{2}} e^{ - x}$, which leads to

\[
\mathop {\lim }\limits_{b \to \infty } \left( {b - a} \right)^{ - \lambda _0 ^2 ab\frac{{b + a}}{{b - a}}} \sqrt {\frac{{\Gamma \left( {n + 2\lambda _0 ^2 ab\frac{{b + a}}{{b - a}} + 1} \right)}}{{\Gamma \left( {n + 2\lambda _0 ^2 \frac{{ab^2 }}{{b - a}} + 1} \right)}}} \left( {b - x} \right)^{\lambda _0 ^2 \frac{{ab^2 }}{{b - a}}}  = \left( {2\lambda _0 ^2 a} \right)^{\lambda _0 ^2 a^2 } e^{ - \lambda _0 ^2 a\left( {x - a} \right)} .
\]

Concluding, we want to highlight that a special approach to the quantum well problem is used for the construction of its two unique models. These models are completely defined within the positive position representation. They behave as oscillator-shaped, however, their mass is not constant, but varies with position. This feature of the mass allows for considering these models as confined quantum wells. Also, if one attempts to extend the definition of these models to the negative position values, then an interesting singularity arises changing the mass from positive to the negative one. Such negativity of the mass also changes potential itself from a positive to a negative definition. At present, this point of discussions is out of the aim of the paper, however, in the future, this property can be the attractive initial point of discussion regarding exactly solvable models of the inverted harmonic oscillator problem (cf. with~\cite{kalnins1974,rotbart1978,caldeira1981,fertig1987,baskoutas1994,dattagupta1997,yuce2006,sanin2007,munoz2009,bermudez2013,pachon2014,kumar2014,maamche2017,subramanyan2021}, where similar oscillator problem is discussed via replacement $\omega \to i \omega$ in the potential energy).

\section*{Acknowledgement}

This work was supported by the Azerbaijan Science Foundation through Grant No. \textbf{AEF-MCG-2022-1(42)-12/01/1-M-01}.

\end{document}